# Scalable Privacy-Preserving Data Sharing Methodology for Genome-Wide Association Studies

Fei Yu

*Department of Statistics, Carnegie Mellon University, Pittsburgh, PA 15213-3890, USA*

Stephen E. Fienberg

*Department of Statistics, Heinz College, Machine Learning Department, and Cylab, Carnegie Mellon University, Pittsburgh, PA 15213-3890, USA*

Aleksandra Slavković

*Department of Statistics, Department of Public Health Sciences, Penn State University, University Park, PA 16802 USA*

Caroline Uhler

*Institute of Science and Technology Austria, Am Campus 1, 3400 Klosterneuburg, Austria*

**Abstract**

The protection of privacy of individual-level information in genome-wide association study (GWAS) databases has been a major concern of researchers following the publication of "an attack" on GWAS data by Homer et al. [1]. Traditional statistical methods for confidentiality and privacy protection of statistical databases do not scale well to deal with GWAS data, especially in terms of guarantees regarding protection from linkage to external information. The more recent concept of differential privacy, introduced by the cryptographic community, is an approach that provides a rigorous definition of privacy with meaningful privacy guarantees in the presence of arbitrary external information, although the guarantees may come at a serious price in terms of data utility. Building on such notions, Uhler et al. [2] proposed

*Email addresses:* `feiy@stat.cmu.edu` (Fei Yu), `fienberg@stat.cmu.edu` (Stephen E. Fienberg), `sesa@psu.edu` (Aleksandra Slavković), `caroline.uhler@ist.ac.at` (Caroline Uhler)

new methods to release aggregate GWAS data without compromising an individual's privacy. We extend the methods developed in [2] for releasing differentially-private $\chi^2$-statistics by allowing for arbitrary number of cases and controls, and for releasing differentially-private allelic test statistics. We also provide a new interpretation by assuming the controls' data are known, which is a realistic assumption because some GWAS use publicly available data as controls. We assess the performance of the proposed methods through a risk-utility analysis on a real data set consisting of DNA samples collected by the Wellcome Trust Case Control Consortium and compare the methods with the differentially-private release mechanism proposed by Johnson and Shmatikov [3].

## 1. Introduction

A genome-wide association study (GWAS) tries to identify genetic variations that are associated with a disease. A typical GWAS examines single-nucleotide polymorphisms (SNPs) from thousands of individuals and produces aggregate statistics, such as the $\chi^2$-statistic and the corresponding *p*-value, to evaluate the association of a SNP with a disease.

For many years researchers have assumed that it is safe to publish aggregate statistics of SNPs that they found most relevant to the disease. Because these aggregate statistics were pooled from thousands of individuals, they believed that their release would not compromise the participants' privacy. However, such belief was challenged when Homer et al. [1] demonstrated that, under certain conditions, given an individual's genotype, one only needs the minor allele frequencies (MAFs) in a study and other publicly available MAF information, such as SNP data from the HapMap[1] project, in order to "accurately and robustly" determine whether the individual is in the test population or the reference population. Here, the test population can be the cases in a study, and the reference population can be the data from the HapMap project. Homer et al. [1] defined a distance metric that contrasts the similarity between an individual and the test population and that between the individual and the reference population, and constructed a *t*-test based on this distance metric. They then showed that their method of identifying an individual's membership status has almost zero false positive rate and zero false negative rate.

---

[1]http://hapmap.ncbi.nlm.nih.gov/

However, Braun et al. [4] argued that the key assumptions of the Homer et al. [1] attack are too stringent to be applicable in realistic settings. Most problematic are the assumptions that (i) the SNPs are in linkage equilibrium and (ii) that the individual, the reference population, and the test population are samples from the same underlying population. They presented a sensitivity analysis of the key assumptions and showed that violation of the first assumption results in a substantial increase in variance and violation of the second condition, together with the condition that the reference population and the test population have different sizes, results in the test statistic deviating considerably from the standard normal distribution.

Notwithstanding the apparent limitation of the Homer et al. [1] attack, the National Institute of Health (NIH) was cautious about the potential breach of privacy in genetic studies (see Couzin [5] and Zerhouni and Nabel [6]), and swiftly instituted an elaborate approval process that every researcher has to go through in order to gain access to aggregate genetic data.[2,3] This NIH policy remains in effect today.

The paper by Homer et al. [1] attracted considerable attention within the genetics community and spurred interest in investigating the vulnerability of confidentiality protection of GWAS databases. The research efforts include modifications and extensions of the Homer et al. attack, alternative formulations of the identification problem, and different aspects of attacking and protecting the GWAS databases; e.g., see [7–17]. In partial response to this literature, Uhler et al. [2] proposed new methods for releasing aggregate GWAS data without compromising an individual's privacy by focusing on the release of differentially-private minor allele frequencies, $\chi^2$-statistics and *p*-values.

In this paper, we develop a differentially-private allelic test statistic and extend the results on differentially-private $\chi^2$-statistics in [2] to allow for an arbitrary number of cases and controls. We start with some main definitions and notation in Section 2. The new sensitivity results are presented in

---

[2]http://gwas.nih.gov/pdf/Data%20Sharing%20Policy%20Modifications.pdf
[3]http://epi.grants.cancer.gov/dac/da_request.html

Section 3. Uhler et al. [2] proposed an algorithm based on the *Laplace mechanism* for releasing the $M$ most relevant SNPs in a differentially-private way. In the same paper they also developed an alternative approach to differential privacy in the GWAS setting using what is known as the *exponential mechanism* linked to an objective function perturbation method by Chaudhuri et al. [18]. This was proposed as a way to achieve a differentially-private algorithm for detecting epistasis. But the *exponential mechanism* could in principle have also been used as a direct alternative to the *Laplace mechanism* of Uhler et al. [2]. This is in fact what Johnson and Shmatikov [3] proposed. Their method selects the top-ranked $M$ SNPs using the exponential mechanism. In Section 4 we review the algorithm based on the Laplace mechanism from [2] and propose a new algorithm based on the exponential mechanism by adapting the method by Johnson and Shmatikov [3]. Finally, in Section 5 we compare our two algorithms to the algorithm proposed in [3] by analyzing a data set consisting of DNA samples collected by the Wellcome Trust Consortium (WTCCC)[4] and made available to us for reanalysis.

## 2. Main Definitions and Notation

The concept of differential privacy, recently introduced by the cryptographic community (e.g., Dwork et al. [19]), provides a notion of privacy guarantees that protect GWAS databases against arbitrary external information.

**Definition 1.** Let $\mathcal{D}$ denote the set of all data sets. Write $D \sim D'$ if $D$ and $D'$ differ in one individual. A randomized mechanism $\mathcal{K}$ is $\epsilon$-differentially private if, for all $D \sim D'$ and for any measurable set $S \subset \mathbb{R}$,

$$\frac{Pr(\mathcal{K}(D) \in S)}{Pr(\mathcal{K}(D') \in S)} \leq e^\epsilon.$$

**Definition 2.** The sensitivity of a function $f : \mathcal{D}^N \to \mathbb{R}^d$, where $\mathcal{D}^N$ denotes the set of all databases with $N$ individuals, is the smallest number $S(f)$ such that

$$||f(D) - f(D')||_1 \leq S(f),$$

for all data sets $D, D' \in \mathcal{D}^N$ such that $D \sim D'$.

---

[4] http://www.wtccc.org.uk/

Table 1: Genotype distribution

|  | # of minor alleles | | | |
|---|---|---|---|---|
|  | 0 | 1 | 2 | Total |
| Case | $r_0$ | $r_1$ | $r_2$ | R |
| Control | $s_0$ | $s_1$ | $s_2$ | S |
| Total | $n_0$ | $n_1$ | $n_2$ | N |

Table 2: Allelic distribution

|  | Allele type | | |
|---|---|---|---|
|  | Minor | Major | Total |
| Case | $r_1 + 2r_2$ | $2r_0 + r_1$ | 2R |
| Control | $s_1 + 2s_2$ | $2s_0 + s_1$ | 2S |
| Total | $n_1 + 2n_2$ | $2n_0 + n_1$ | 2N |

Releasing $f(D) + b$, where $b \sim \text{Laplace}\left(0, \frac{S(f)}{\epsilon}\right)$, satisfies the definition of $\epsilon$-differential privacy (e.g., see [19]). This type of release mechanism is often referred to as the *Laplace mechanism*. Here $\epsilon$ is the privacy budget; a smaller value of $\epsilon$ implies stronger privacy guarantees.

*2.1. SNP Summaries Using Contingency Tables*

Following the notation in [20], we can summarize the data for a single SNP in a case-control study with $R$ cases and $S$ controls using a $2 \times 3$ genotype contingency table shown in Table 1, or a $2 \times 2$ allelic contingency table shown in Table 2. We require that margins of the contingency table be positive.

**Definition 3.** The (Pearson) $\chi^2$-statistic based on a genotype contingency table (Table 1) is

$$Y = \frac{(r_0 N - n_0 R)^2}{n_0 RS} + \frac{(r_1 N - n_1 R)^2}{n_1 RS} + \frac{(r_2 N - n_2 R)^2}{n_2 RS}.$$

**Definition 4.** The allelic test is also known as the Cochran-Armitage trend test for the additive model. The allelic test statistic based on a genotype contingency table (Table 1) is equivalent to the $\chi^2$-statistic based on the corresponding allelic contingency table (Table 2). The allelic test statistic can be written as

$$Y_A = \frac{2N^3}{RS} \frac{\{(s_1 + 2s_2) - \frac{S}{N}(n_1 + 2n_2)\}^2}{2N(n_1 + 2n_2) - (n_1 + 2n_2)^2}.$$

The Pearson $\chi^2$-test for genotype data and the allelic test for allele data are among the most commonly used statistical tests for association in GWAS. Zheng et al. [21] suggest using the allelic test when the genetic model of the phenotype is additive, and the Pearson $\chi^2$-test when the genetic model is unknown.

## 3. Sensitivity Results

Under the assumption that there are an equal number of cases and controls, Uhler et al. [2] found the sensitivities of the $\chi^2$-statistic, the corresponding $p$-value and the projected $p$-value. For completeness, we briefly review these results here.

**Theorem 3.1** (Uhler et al. [2])**.** *The sensitivity of the $\chi^2$-statistic based on a $3\times 2$ contingency table with positive margins and $N/2$ cases and $N/2$ controls is $\frac{4N}{N+2}$.*

**Theorem 3.2** (Uhler et al. [2])**.** *The sensitivity of the p-values of the $\chi^2$-statistic based on a $3 \times 2$ contingency table with positive margins and $N/2$ cases and $N/2$ controls is $\exp(-2/3)$, when the null distribution is a $\chi^2$-distribution with 2 degrees of freedom.*

**Corollary 3.3** (Uhler et al. [2])**.** *Projecting all p-values larger than $p^* = \exp(-N/c)$ onto $p^*$ results in a sensitivity of $\exp(-N/c)-\exp\left(-\frac{N(2Nc-4N-4c+c^2)}{2c(Nc-2N-c)}\right)$ for any fixed constant $c \geq 3$, which is a factor of $N/2$.*

In the remainder of this section, we generalize these results to allow for an arbitrary number of cases and controls. This makes the proposed methods applicable in a typical GWAS setting, in which there are more controls than cases, as researchers often use data pertaining to other diseases as controls to increase the statistical power.

*3.1. Sensitivity Results for the Pearson $\chi^2$-Statistic*

We first consider the situation in which the adversary has complete information about the controls. This situation arises when a GWAS uses publicly available data for the controls, such as those from the HapMap project. In this scenario, it is only necessary to protect information about the cases.

**Theorem 3.4.** *Let $\mathcal{D}$ denote the set of all $2 \times 3$ contingency tables with positive margins, $R$ cases and $S$ controls. Suppose the numbers of controls of all three genotypes are known. Let $N = R + S$, and $s_{\max} = \max\{s_0, s_1, s_2\}$. The sensitivity of the $\chi^2$-statistic based on tables in $\mathcal{D}$ is bounded above by $\frac{N^2}{RS} \frac{s_{\max}}{1+s_{\max}}$.*

*Proof.* See Appendix A. □

Theorem 3.4 gives an upper bound for the sensitivity of the $\chi^2$-statistic based on $2 \times 3$ contingency tables with positive margins and known numbers of controls for all three genotypes. In Corollary 3.5 we show that, assuming $r_0 \geq r_2$ and $s_0 \geq s_2$, which reflects the definition of a major and minor allele, the upper bound for the sensitivity is attained.

**Corollary 3.5.** *Let $\mathcal{D}$ denote the set of all $2 \times 3$ contingency tables with positive margins, $R$ cases and $S$ controls. We further assume that for tables in $\mathcal{D}$, $r_0 \geq r_2$ and $s_0 \geq s_2$; i.e., in the case and control populations the number of individuals having two minor alleles is no greater than the number of individuals having two major alleles. The sensitivity of the $\chi^2$-statistic based on tables in $\mathcal{D}$ is $\frac{N^2}{RS}\left(1 - \frac{1}{\max\{S,R\}+1}\right)$, where $N = R + S$.*

*Proof.* For a change that occurs in the cases, we first treat $s_0$, $s_1$, and $s_2$ as fixed, and get the result in Theorem 3.4. By taking $(r_0, r_1, r_2, s_0, s_1, s_2) = (r_0, 1, r_2, 0, S, 0)$, $r_0 \geq r_2 > 0$, and changing the table in the direction of $u = (1, -1, 0, 0, 0, 0)$, we attain the upper bound $\frac{N^2}{RS}\left(1 - \frac{1}{S+1}\right)$. The same analysis for a change that occurs in the controls shows that the maximum change of the Pearson $\chi^2$-statistic (i.e., $Y$ in Appendix A) is $\frac{N^2}{RS}\left(1 - \frac{1}{R+1}\right)$. □

If we have no knowledge of either the cases or the controls, we get the sensitivity result presented in Corollary 3.5. On the other hand, when the controls are known, we can use Theorem 3.4 to reduce the sensitivity assigned to each set of SNPs grouped by the maximum number of controls among the three genotypes. However, in most GWAS the number of controls, $S$, is large and $s_{\max} = \max\{s_0, s_1, s_2\} \geq S/3$. In this case, the following computation shows that the reduction in sensitivity obtained by Theorem 3.4 is

insignificant:

$$\left\{\frac{N^2}{RS}\frac{S}{1+S}\right\}\bigg/\left\{\frac{N^2}{RS}\frac{s_{\max}}{1+s_{\max}}\right\} \leq \left\{\frac{N^2}{RS}\frac{S}{1+S}\right\}\bigg/\left\{\frac{N^2}{RS}\frac{S/3}{1+S/3}\right\}$$
$$= \frac{S+3}{S+1}$$
$$\approx 1.$$

In order to improve on statistical utility, Uhler et al. [2] proposed projecting the $p$-values that are larger than a threshold value onto the threshold value itself to reduce the sensitivity. In Theorem 3.6 we generalize this result to nonnegative score functions, showing how to incorporate projections into the Laplace mechanism.

**Theorem 3.6.** *Given a nonnegative function $f(d)$, define $h_C(d) = \max\{C, f(d)\}$, with $C > 0$; i.e., we project values of $f(d)$ that are smaller than $C$ onto $C$. Let $s$ denote the sensitivity of $h_C(d)$, and suppose $Y \sim \text{Laplace}(0, \frac{s}{\epsilon})$, then $W(d) = \max\{C, Z(d)\}$, with $Z(d) = h_C(d) + Y$, is $\epsilon$-differentially private.*

*Proof.* From the definition of $W(d)$, we know that $W(d) \geq C$ for all $d$. For $t > C$,

$$\frac{\mathrm{P}(W(d) = t)}{\mathrm{P}(W(d') = t)} = \frac{\mathrm{P}(Z(d) = t)}{\mathrm{P}(Z(d') = t)} \leq \exp\left(\left||t - h_C(d')| - |t - h_C(d)|\right|\epsilon/s\right)$$
$$\leq \exp\left(|h_C(d) - h_C(d')|\epsilon/s\right)$$
$$\leq \exp(\epsilon).$$

For $t = C$,

$$\frac{\mathrm{P}(W(d) = C)}{\mathrm{P}(W(d') = C)} = \frac{\mathrm{P}(Z(d) \leq C)}{\mathrm{P}(Z(d') \leq C)} = \frac{\frac{1}{2}\exp\left(\frac{C - h_C(d)}{s/\epsilon}\right)}{\frac{1}{2}\exp\left(\frac{C - h_C(d')}{s/\epsilon}\right)}$$
$$\leq \exp\left(|h_C(d') - h_C(d)|\epsilon/s\right)$$
$$\leq \exp(\epsilon).$$

$\square$

For example, when we apply this result to $\chi^2$-statistics in a differentially-private mechanism, we set $C$ to be the $\chi^2$-statistic that corresponds to a small $p$-value and use an upper bound for the sensitivity of the projection function as $s_C$, namely

$$s_C = \min\left\{Y_{\max} - C, \frac{N^2}{RS}\left(1 - \frac{1}{\max\{S, R\} + 1}\right)\right\}.$$

*3.2. Sensitivity Results for the Allelic Test Statistic*

**Theorem 3.7.** *The sensitivity of the allelic test statistic based on a $2 \times 3$ contingency table with positive margins, $R$ cases and $S$ controls is given by the maximum of*

$$\left\{\begin{array}{l} \dfrac{8N^2 S}{R(2S+3)(2S+1)}, \\ \dfrac{4N^2[(2R^2-1)(2S-1)-1]}{RS(2R+1)(2R-1)(2S+1)}, \\ \dfrac{8N^2 R}{S(2R+3)(2R+1)}, \\ \dfrac{4N^2[(2S^2-1)(2R-1)-1]}{RS(2S+1)(2S-1)(2R+1)} \end{array}\right\}.$$

*Proof.* See Appendix B. □

## 4. Privacy-Preserving Release of the Top $M$ Statistics

In a GWAS setting, researchers usually assign to every SNP a score that reflects its association with a disease, but only release scores for the $M$ most significant SNPs. Most commonly used scores are the Pearson $\chi^2$-statistic, the allelic test statistic, and the corresponding $p$-values. If those $M$ SNPs were chosen according to a uniform distribution, $\epsilon$-differential privacy can be achieved by the Laplace mechanism with noise $\frac{Ms}{\epsilon}$, where $s$ denotes the sensitivity of the scoring statistic. Recall that $\epsilon$ is the privacy budget, so a smaller value of $\epsilon$ implies stronger privacy guarantees.

However, by releasing $M$ SNPs according to their rankings, an attacker knows that the released SNPs have higher scores than all other SNPs regardless of the face value of the released scores. Therefore, we need a more sophisticated algorithm for releasing the $M$ most significant SNPs.

**Algorithm 1** The $\epsilon$-differentially private algorithm for releasing the $M$ most relevant SNPs using the *Laplace mechanism*.

> **Input:** The score (e.g., $\chi^2$-statistic or allelic test statistic) used to rank all $M'$ SNPs, the number of SNPs, $M$, that we want to release, the sensitivity, $s$, of the statistic, and $\epsilon$, the privacy budget.
>
> **Output:** $M$ noisy statistics.
>
> 1. Add Laplace noise with mean zero and scale $\frac{4Ms}{\epsilon}$ to the true statistics.
> 2. Pick the top $M$ SNPs with respect to the perturbed statistics. Denote the corresponding set of SNPs by $\mathcal{S}$.
> 3. Add new Laplace noise with mean zero and scale $\frac{2Ms}{\epsilon}$ to the true statistics in $\mathcal{S}$.

Adapting from the differentially-private algorithm for releasing the most frequent patterns in Bhaskar et al. [22], Uhler et al. [2] suggested an algorithm (see Algorithm 1) for releasing the $M$ most relevant SNPs ranked by their $\chi^2$-statistics or the corresponding $p$-values while satisfying differential privacy. They also showed that adding noise directly to the $\chi^2$-statistic achieves a better trade-off between privacy and utility than by adding noise to the $p$-values or cell entries themselves. Using the results from Section 3, we can now also apply this algorithm when the number of cases and controls differ.

While Algorithm 1 is based on the Laplace mechanism, in Algorithm 2 we propose a new algorithm based on the exponential mechanism by adopting and simplifying the ideas proposed by Johnson and Shmatikov [3]. The first application of the exponential mechanism in the GWAS setting was given in [2], which resulted in a differentially private algorithm for detecting epistasis.

**Theorem 4.1.** *Algorithm 2 is $\epsilon$-differentially private.*

*Proof.* See Appendix C. □

## 5. Application of Differentially Private Release Mechanisms to Human GWAS Data

In this section we evaluate the trade-off between data utility and privacy risk by applying Algorithm 1 and Algorithm 2 with the new sensitivity results developed in Section 3 to a GWAS data set containing human DNA samples from WTCCC. We also compare the performance of Algorithm 1

**Algorithm 2** The $\epsilon$-differentially private algorithm for releasing the $M$ most relevant SNPs using the *exponential mechanism*.

> **Input:** The score (e.g., $\chi^2$-statistic or allelic test statistic) used to rank all $M'$ SNPs, the number of SNPs, $M$, that we want to release, the sensitivity, $s$, of the statistic, and $\epsilon$, the privacy budget.
>
> **Output:** $M$ noisy statistics.
>
> 1. Let $\mathcal{S} = \emptyset$ and $q_i =$ score of $\text{SNP}_i$.
> 2. For $i \in \{1, \ldots, M'\}$, set $w_i = \exp\left(\frac{\epsilon q_i}{4Ms}\right)$.
> 3. Set $p_i = w_i \left/ \sum_{j=1}^{M'} w_j \right., i \in \{1, \ldots, M'\}$, the probability of sampling the $i$th SNP.
> 4. Sample $k \in \{1, \ldots, M'\}$ with probability $\{p_1, \ldots, p_{M'}\}$. Add $\text{SNP}_k$ to $\mathcal{S}$. Set $q_k = -\infty$.
> 5. If the size of $\mathcal{S}$ is less than $M$, return to Step 2.
> 6. Add new Laplace noise with mean zero and scale $\frac{2Ms}{\epsilon}$ to the true statistics in $\mathcal{S}$.

and Algorithm 2 to that of the `LocSig` method developed by Johnson and Shmatikov [3]. Essential to the `LocSig` method is a scoring function based on the $p$-value of a statistical test. In this paper, we call the resulting scores the *JS scores*. In contrast to [3], which used the $p$-value of the G-test to construct the JS scores, we use the $p$-value of the Pearson $\chi^2$-test instead.

*5.1. Data Set from WTCCC: Crohn's Disease*

We use a real data set that was collected by the WTCCC and intended for genome-wide association studies of Crohn's disease. The data set consists of DNA samples from 3 cohorts, the subjects of which all lived within Great Britain and identified themselves as white Europeans: *1958 British Birth Cohort* (58C), *UK Blood Services* (NBS), and *Crohn's disease* (CD). In the original study [23] the DNA samples from the 58C and NBS cohorts are treated as controls and those from the CD cohort as cases.

The data were sampled using the Affymetrix GeneChip 500K Mapping Array Set. The genotype data were called by an algorithm named CHI-AMO (see [23]), which WTCCC developed and deemed more powerful than Affymetrix's BRLMM genotype calling algorithm. According to the WTCCC analysis, some DNA samples were contaminated or came from non-Caucasian ancestry. In addition, they indicated that some SNPs did not pass quality control filters. Finally, WTCCC [23] removed additional SNPs from their analysis by visually inspecting cluster plots.

5.2. Our Re-Analysis of the WTCCC Data

In [23], the authors mainly used the allelic test and the Pearson $\chi^2$-test to find SNPs with a strong association with Crohn's disease, and reported the relevant statistics and their $p$-values for the most significant SNPs. In general, the Wellcome Trust Case Control Consortium [23] considered a SNP significant if its allelic test $p$-value or $\chi^2$-test $p$-value were smaller than $10^{-5}$. In the supplementary material of [23] they reported 26 significant SNPs, 6 of which were imputed. Per [23], imputing SNPs that do not exist in the WTCCC databases does not affect the calculation of the allelic test statistics or the Pearson $\chi^2$-statistics of SNPs already in the WTCCC databases; therefore, we disregard the imputed SNPs in our analysis and retain 20 significant SNPs.

We followed the filtering process in [23] closely and removed DNA samples and SNPs that [23] deemed contaminated. However, we did not remove any further SNPs due to poor cluster plots. We verified that our processing of the raw genotype data leads to the same results as those published in the supplementary material of [23]: our calculations for 16 of the 20 reported significant SNPs match those in [23], deviating no more than 2% in allelic test statistic and $\chi^2$-statistic. However, we found that a number of significant SNPs were not reported by the WTCCC. We corresponded with one of the principal authors of [23] and received confirmation that the WTCCC also found those SNPs to be significant. However, Wellcome Trust Case Control Consortium [23] did not report these SNPs because they suffered from poor calling quality according to visual inspection of the cluster plot, a procedure that we did not implement. We excluded from our analysis these SNPs that have significant allelic test $p$-values or $\chi^2$-test $p$-values, but are not reported by the WTCCC.

In Figure 1 we plot the $\chi^2$-statistics resulting from our analysis in descending order. Note that there is a large gap between the 5th and the 6th largest $\chi^2$-statistics. This is an important observation for the risk-utility analysis of the perturbed statistics in Section 5.3. Because of the nature of the distribution of the top $\chi^2$-statistics in this data set, it is easier to recover all top 5 SNPs as the top rated 5 SNPs in the perturbed data than it is to recover all top $M$ SNPs for $M < 5$ or $M > 5$, as is evident in Figure 2, which we discuss in the next section.

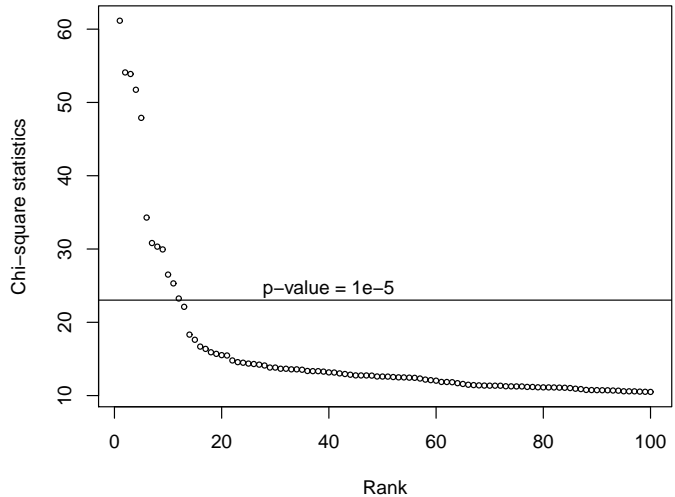

Figure 1: Unperturbed top χ-statistics in descending order.

To summarize, we were able to reproduce a high percentage of significant SNPs from [23]. Therefore, we are confident that our data processing procedure is sound and the $\chi^2$-statistics and allelic test statistics that we obtained from the data are comparable with those produced in a high quality GWAS.

*5.3. Risk-Utility Analysis of Differentially Private Pearson $\chi^2$-statistics*

In this section, we use the $\chi^2$-statistics obtained from the WTCCC dataset described in Section 5.1 and analyze the statistical utility of releasing differentially private $\chi^2$-statistics for various privacy budgets, $\epsilon$. With 1748 cases and 2938 controls in the WTCCC dataset, we use Corollary 3.5 to obtain a sensitivity of 4.27 for the $\chi^2$-statistic.

We define statistical utility as follows: let $\mathcal{S}_0$ be the set of top $M$ SNPs ordered according to their true $\chi^2$-statistics and let $\mathcal{S}$ be the set of top $M$ SNPs chosen after perturbation (either by Algorithm 1 using the Laplace mechanism or by Algorithm 2 using the exponential mechanism). Then the utility as a function of $\epsilon$ is

$$u(\epsilon) = \frac{|\mathcal{S}_0 \cap \mathcal{S}|}{|\mathcal{S}_0|}.$$

We perform the following procedure to approximate the expected utility $\mathbb{E}[u(\epsilon)]$ for Algorithm 1: (i) add Laplace noise with mean zero and scale $\frac{4Ms}{\epsilon}$ to the true $\chi^2$-statistics, where $s$ is the sensitivity of the $\chi^2$-statistic; (ii) pick the top $M$ SNPs with respect to the perturbed $\chi^2$-statistics; (iii) denote the set of SNPs chosen according to the true and perturbed $\chi^2$-statistics by $\mathcal{S}_0$ and $\mathcal{S}$, respectively; (iv) calculate $u(\epsilon) = \frac{|\mathcal{S}_0 \cap \mathcal{S}|}{|\mathcal{S}_0|}$. We repeat the aforementioned procedure 50 times for a fixed $\epsilon$ and report the average of the utility $u(\epsilon)$. To approximate $\mathbb{E}[u(\epsilon)]$ for Algorithm 2, we repeat 50 times the process of generating $\mathcal{S}$ by performing steps 1–5 in Algorithm 2 and report the average of the utility $u(\epsilon)$. In order to approximate $\mathbb{E}[u(\epsilon)]$ for the procedure `LocSig` from [3], we rank the SNPs by their $\chi^2$-statistics but replace the scores in Step 1 by the JS scores.

The runtimes of the different algorithms vary considerably (see Table 3). The runtimes were obtained on a PC with an Intel i5-3570K CPU, 32 GB of RAM and the Ubuntu 13.04 operating system. Calculating the $\chi^2$-statistics from genotype tables is a trivial task and takes very little time. Calculating the JS scores can be a daunting task, however, if one cannot find a clever simplification. The JS score is essentially the shortest Hamming distance between the original database and the set of databases at which the significance of the $p$-value changes. Thus without any simplifications, one would need to search the entire space of databases in order to find the table with the shortest Hamming distance. In our implementation for finding the JS score for a genotype table based on the $p$-value corresponding to the $\chi^2$-statistic, we simplify the calculation by greedily following the path of maximum change of the $\chi^2$-statistic until we find a table with altered

Table 3: Comparison of runtime for the simulations in Section.5.3 The number of repetitions is 50, the number of different values for $M$ is 4, the number of different values for $\epsilon$ is 15, and the number of SNPs is around 4000. $\mathcal{S}$ is the set of SNPs to be released after the perturbation.

| Method | Time spent on generating $\mathcal{S}$ (in minutes) | Time spent on calculating the scores (in minutes) |
|---|---|---|
| Algorithm 1 (Laplace) | 0.04 | $\approx 0$ |
| Algorithm 2 (Exponential) | 1.53 | $\approx 0$ |
| `LocSig` (JS) | 2.00 | 3.50 |

significance.

In Figure 2, we compare the performance of Algorithm 1 (based on the Laplace mechanism) and Algorithm 2 (based on the exponential mechanism) to `LocSig` ([3]). It is clear that when $\epsilon = 1$, the `LocSig` method outperforms the other methods with respect to utility. Nevertheless, we note a few features regarding the performance of `LocSig` .

- When $M = 3$, the utility of the `LocSig` method cannot exceed 0.67 even as $\epsilon$ continues to increase. This artifact is due to the fact that the ranking of SNPs based on the JS scores is different from the ranking based on the $\chi^2$-statistics.

- Table 4 gives the top 6 SNPs ranked by their $\chi^2$-statistics and the corresponding JS scores. For all threshold $p$-values, the JS score of the 4th SNP is larger than that of the 2nd SNP and that of the $i$th SNP for $i \geq 5$. Thus, when $\epsilon$ is sufficiently large, the `LocSig` method will almost always output the 1st, 3rd, and 4th SNPs. Consequently, the utility for the `LocSig` method will not increase when $\epsilon$ increases.

- The `LocSig` method is sensitive to the choice of $p$-value. This becomes apparent in the plots for $M = 15$ in Figure 2. The risk-utility curves of the `LocSig` method tend to have lower utility for the same $\epsilon$ when the threshold $p$-value is smaller.

- Even though Algorithm 1 and Algorithm 2 do not perform as well as the `LocSig` method for small values of $\epsilon$, they do not suffer from the aforementioned issues. Furthermore, we can see from Figure 2 that the exponential mechanism always outperforms the Laplace mechanism, i.e., it achieves a higher utility for each value of $\epsilon$.

Table 4: Ranking of the top 6 SNPs by $\chi^2$-statistics and the corresponding JS scores. $K$ denotes the total number of SNPs.

| Scoring scheme | Threshold $p$-value | Score (nearest integer) | | | | | |
|---|---|---|---|---|---|---|---|
| | | 1st | 2nd | 3rd | 4th | 5th | 6th |
| $\chi^2$-statistic | - | 61 | 54 | 54 | 52 | 48 | 34 |
| JS score | $0.001/K$ | 51 | 31 | 37 | 33 | 25 | 6 |
| JS score | $0.01/K$ | 61 | 38 | 47 | 41 | 33 | 13 |
| JS score | $0.05/K$ | 69 | 43 | 55 | 48 | 38 | 18 |

To summarize, in this application Algorithm 2 outperforms Algorithm 1. The method based on `LocSig` improves on Algorithm 2 for small values of $\epsilon$, but shows some problematic behavior when $\epsilon$ increases. Finally, the `LocSig` method comes at a much higher computational cost than the other two algorithms and might not be computationally feasible for some data sets.

## 6. Conclusions

A number of authors have argued that it is possible to use aggregate data to compromise the privacy of individual-level information collected in GWAS databases. We have used the concept of differential privacy and built on the approach in Uhler et al. [2] to propose new methods to release aggregate GWAS data without compromising an individual's privacy. A key component of the differential privacy approach involves the sensitivity of a released statistic when we remove an observation. In this paper, we have obtained sensitivity results for the Pearson $\chi^2$-statistic when there are arbitrary number of cases and controls. Furthermore, we showed that the sensitivity can be reduced in the situation where data for the cases (or the controls) are known to the attacker. Nevertheless, we also showed that the reduction in sensitivity is insignificant in typical GWAS, in which the number of cases is large.

By incorporating the two-step differentially-private mechanism for releasing the top $M$ SNPs (Algorithm 1) with the projected Laplace perturbation mechanism (Theorem 3.6), we have created an algorithm that outputs significant SNPs while preserving differential privacy. We demonstrated that the algorithm works effectively in human GWAS datasets, and that it produces outputs that resemble the outputs of regular GWAS. We also showed that the performance of Algorithm 1, which is based on the Laplace mechanism, can be improved by using Algorithm 2, which is based on the exponential mechanism. Furthermore, Algorithm 2 is computationally more efficient than the `LocSig` method of Johnson and Shmatikov [3], and it performs better for increasing values of $\epsilon$.

Finally, we showed that a risk-utility analysis of the algorithm allows us to understand the trade-off between privacy budget and statistical utility, and therefore helps us decide on the appropriate level of privacy guarantee for the released data. We hope that approaches such as those that we demonstrate in this paper will allow the release of more information from GWAS going forward and allay the privacy concerns that others have voiced over the past decade.

Figure 2: Performance comparison of Algorithm 1 ("Laplace"), Algorithm 2 with $\chi^2$-statistics as scores ("Exponential"), and the `LocSig` method in Johnson and Shmatikov [3] ("JS") based on the $p$-value of the $\chi^2$-statistic. Each row corresponds to a fixed $M$, the number of top SNPs to release. Each column corresponds to a fixed threshold $p$-value, which is relevant to the `LocSig` method only; it is irrelevant to the other methods. Data used to generate this figure consist of SNPs with $p$-values smaller than $10^{-5}$ and a randomly chosen 1% sample of SNPs with $p$-values larger than $10^{-5}$; the total number of SNPs used for calculation is 3882.

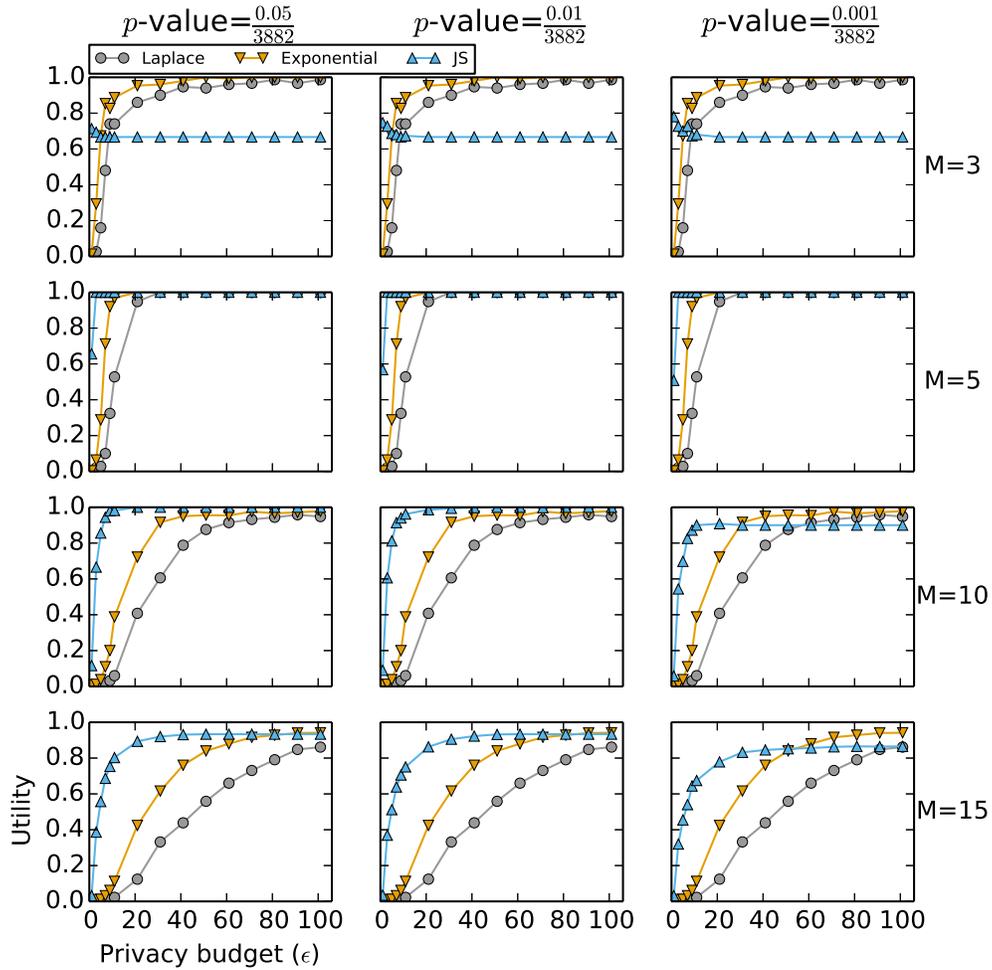

## 7. Acknowledgment

This research was partially supported by NSF Awards EMSW21-RTG and BCS-0941518 to the Department of Statistics at Carnegie Mellon University, and by NSF Grant BCS-0941553 to the Department of Statistics at Pennsylvania State University. This work was also supported in part by the National Center for Research Resources, Grant UL1 RR033184, and is now at the National Center for Advancing Translational Sciences, Grant UL1 TR000127 to Pennsylvania State University. The content is solely the responsibility of the authors and does not necessarily represent the official views of the NSF and NIH. We thank the Wellcome Trust Consortium for giving us access to the data used in this paper to illustrate our methods.

# Appendices

### Appendix A. Proof of Theorem 3.4

*Proof.* The Pearson $\chi^2$-statistic can be written as

$$
\begin{aligned}
Y &= \frac{\left(r_0 - \frac{n_0 R}{N}\right)^2}{\frac{n_0 R}{N}} + \frac{\left(r_1 - \frac{n_1 R}{N}\right)^2}{\frac{n_1 R}{N}} + \frac{\left(r_2 - \frac{n_2 R}{N}\right)^2}{\frac{n_2 R}{N}} \\
&\quad + \frac{\left(s_0 - \frac{n_0 S}{N}\right)^2}{\frac{n_0 S}{N}} + \frac{\left(s_1 - \frac{n_1 S}{N}\right)^2}{\frac{n_1 S}{N}} + \frac{\left(s_2 - \frac{n_2 S}{N}\right)^2}{\frac{n_2 S}{N}} \\
&= (r_0 N - n_0 R)^2 \left(\frac{1}{n_0 RN} + \frac{1}{n_0 SN}\right) \\
&\quad + (r_1 N - n_1 R)^2 \left(\frac{1}{n_1 RN} + \frac{1}{n_1 SN}\right) \\
&\quad + (r_2 N - n_2 R)^2 \left(\frac{1}{n_2 RN} + \frac{1}{n_2 SN}\right) \\
&= \frac{(r_0 N - n_0 R)^2}{n_0 RS} + \frac{(r_1 N - n_1 R)^2}{n_1 RS} + \frac{(r_2 N - n_2 R)^2}{n_2 RS} \\
&= \frac{r_0^2 N^2}{n_0 RS} - \frac{2 r_0 N}{S} + \frac{n_0 R}{S} \\
&\quad + \frac{r_1^2 N^2}{n_1 RS} - \frac{2 r_1 N}{S} + \frac{n_1 R}{S} \\
&\quad + \frac{r_2^2 N^2}{n_2 RS} - \frac{2 r_2 N}{S} + \frac{n_2 R}{S} \\
&= \frac{N^2}{RS}\left(\frac{r_0^2}{n_0} + \frac{r_1^2}{n_1} + \frac{r_2^2}{n_2}\right) - N\frac{R}{S} \quad\text{(A.1a)} \\
&= \frac{N^2}{RS}\left(\frac{s_0^2}{n_0} + \frac{s_1^2}{n_1} + \frac{s_2^2}{n_2}\right) - N\frac{S}{R}. \quad\text{(A.1b)}
\end{aligned}
$$

We denote a contingency table and its column sums by $v = (r_0, r_1, r_2, s_0, s_1, s_2, n_0, n_1, n_2)$. Let $v' = v + u$, with $v'$ and $v$ differing by Hamming distance 1. Finding the sensitivity of $Y$ boils down to finding $v$ and $u$ that maximize $|Y(v) - Y(v+u)|$.

Suppose $r_0 > 0$ and consider $u = (-1, 1, 0, 0, 0, 0, -1, 1, 0)$. As a conse-quence of (A.1b) we find that

$$Y(v) - Y(v+u) = \left[\frac{N^2}{RS}\left(\frac{s_0^2}{n_0} + \frac{s_1^2}{n_1} + \frac{s_2^2}{n_2}\right) - N\frac{S}{R}\right]$$
$$- \left[\frac{N^2}{RS}\left(\frac{s_0^2}{n_0-1} + \frac{s_1^2}{n_1+1} + \frac{s_2^2}{n_2}\right) - N\frac{S}{R}\right]$$
$$= \frac{N^2}{RS}\left[\frac{s_1^2}{n_1(n_1+1)} - \frac{s_0^2}{n_0(n_0-1)}\right].$$

Because $r_0 > 0$, we get that $n_0 = r_0 + s_0 \geq 1 + s_0$, and

$$0 \leq \frac{s_0^2}{n_0(n_0-1)} \leq \frac{s_0}{n_0} \leq \frac{s_0}{1+s_0} \leq \frac{s_{\max}}{1+s_{\max}}.$$

Similarly,

$$0 \leq \frac{s_1^2}{n_1(n_1+1)} \leq \frac{s_1}{n_1+1} \leq \frac{s_1}{1+s_1} \leq \frac{s_{\max}}{1+s_{\max}}.$$

Therefore,

$$\left|\frac{s_1^2}{n_1(n_1+1)} - \frac{s_0^2}{n_0(n_0-1)}\right| \leq \max\left\{\frac{s_1^2}{n_1(n_1+1)}, \frac{s_0^2}{n_0(n_0-1)}\right\} \leq \frac{s_{\max}}{1+s_{\max}}.$$

A similar analysis for all possible directions $u$ and scenarios in which $r_1 > 0$ or $r_2 > 0$ reveals that the sensitivity of $Y$ is bounded above by $\frac{N^2}{RS}\frac{s_{\max}}{1+s_{\max}}$. □

## Appendix B. Proof of Theorem 3.7

*Proof.* We denote a contingency table and its column sums by $v = (r_0, r_1, r_2, s_0, s_1, s_2, n_0, n_1, n_2)$. With the number of cases, R, and the number of controls, S, fixed, we can simply write $v^s = (s_1, s_2, n_1, n_2)$ or $v^r = (r_1, r_2, n_1, n_2)$. Then the allelic test statistic can be written as

$$Y_A(v^s) = \frac{2N^3}{RS}\frac{\{(s_1 + 2s_2) - \frac{S}{N}(n_1 + 2n_2)\}^2}{2N(n_1 + 2n_2) - (n_1 + 2n_2)^2},$$

$$\text{or}\quad Y_A(v^r) = \frac{2N^3}{RS}\frac{\{(r_1 + 2r_2) - \frac{R}{N}(n_1 + 2n_2)\}^2}{2N(n_1 + 2n_2) - (n_1 + 2n_2)^2}.$$

Let $v' = v+u$, with $v'$ and $v$ differing by Hamming distance 1. Finding the sensitivity of $Y_A$ boils down to finding $v$ and $v'$ that maximize $|Y_A(v) - Y_A(v')|$.

This is equivalent to maximizing $|Y_A(v^s) - Y_A(v^s + u^s)|$ and $|Y_A(v^r) - Y_A(v^r + u^r)|$, with $u^s$ and $u^r$ defined as follows:

when $r_0 > 0$,
$$u_1^s = (0, 0, 1, 0) \quad \text{(Case 0} \to \text{Case 1)}$$
$$u_2^s = (0, 0, 0, 1) \quad \text{(Case 0} \to \text{Case 2)}$$

when $s_0 > 0$,
$$u_1^r = (0, 0, 1, 0) \quad \text{(Control 0} \to \text{Control 1)}$$
$$u_2^r = (0, 0, 0, 1) \quad \text{(Control 0} \to \text{Control 2)}.$$

In other words, when $r_0 > 0$, we search for tables that maximize $|\nabla Y_A(v^s) \cdot u_1^s|_{r_0 > 0}$ or $|\nabla Y_A(v^s) \cdot u_2^s|_{r_0 > 0}$; when $s_0 > 0$, we search for tables that maximize $|\nabla Y_A(v^r) \cdot u_1^r|_{s_0 > 0}$ and $|\nabla Y_A(v^r) \cdot u_2^r|_{s_0 > 0}$.

Let's first consider the case $r_0 > 0$. We have $|\nabla Y_A(v^s) \cdot u_1^s| = \left|\frac{\partial}{\partial n_1} Y_A(v^s)\right|$ and $|\nabla Y_A(v^s) \cdot u_2^s| = \left|\frac{\partial}{\partial n_2} Y_A(v^s)\right|$. Denote by

$$\alpha = \frac{2N^3}{S(N-S)},$$
$$C = (s_1 + 2s_2) - \frac{S}{N}(n_1 + 2n_2),$$
$$D = 2N(n_1 + 2n_2) - (n_1 + 2n_2)^2 = (n_1 + 2n_2)(2n_0 + n_1),$$

then

$$Y_A = \alpha \frac{C^2}{D},$$
$$\frac{\partial}{\partial n_1} Y_A(v^s) = \alpha \frac{-2\left[(N - n_1 - 2n_2)C^2 + \frac{S}{N}DC\right]}{D^2} = \alpha \frac{-2}{D^2}\left[(n_0 - n_2)C^2 + \frac{S}{N}DC\right],$$
$$\frac{\partial}{\partial n_2} Y_A(v^s) = 2 \frac{\partial}{\partial n_1} Y_A(v^s).$$

Therefore, tables that maximize $|\nabla Y_A(v^s) \cdot u_1^s|_{r_0 > 0}$ also maximize $|\nabla Y_A(v^s) \cdot u_2^s|_{r_0 > 0}$. Furthermore, for the same table $v^s$, the change of $Y_A(v^s)$ in the direction of $u_2^s$ is no less than that in the direction of $u_1^s$.

Fixing $n_1$ and $n_2$, $\left|\frac{\partial}{\partial n_1} Y_A(v^s)\right|$ depends only on $s_1$ and $s_2$. So maximizing

$\left|\frac{\partial}{\partial n_1}Y_A(v^s)\right|$ is equivalent to maximizing the absolute value of

$$f(s_1, s_2) := (n_0 - n_2)C^2 + \frac{S}{N}DC$$

$$= \frac{S}{N}DC\ \mathbb{I}_{n_0=n_2} + (n_0 - n_2)\left\{\left[C + \frac{SD}{2N(n_0 - n_2)}\right]^2 - \left[\frac{SD}{2N(n_0 - n_2)}\right]^2\right\}\mathbb{I}_{n_0 \neq n_2}$$

$$= \frac{S}{N}DC\ \mathbb{I}_{n_0=n_2} + (n_0 - n_2)\left\{[g(s_1, s_2)]^2 - \left[\frac{SD}{2N(n_0 - n_2)}\right]^2\right\}\mathbb{I}_{n_0 \neq n_2},$$

where $g(s_1, s_2) = C + \frac{SD}{2N(n_0-n_2)} = (s_1 + 2s_2) + \frac{S(n_1+2n_2)^2}{2N(n_0-n_2)}$. Note that the term $D$ does not depend on $s_1$ or $s_2$. There are three scenarios:

(i) when $n_0 = n_2$, $|f(s_1, s_2)| = \frac{S}{N}D|(s_1 + 2s_2) - \frac{S}{N}(n_1 + 2n_2)|$ is maximized when $s_1 + 2s_2$ is minimized or maximized;

(ii) when $n_0 > n_2$, $|f(s_1, s_2)|$ is maximized when $|g(s_1, s_2)|$ is maximized or minimized, which occurs when $s_1 + 2s_2$ is minimized or maximized;

(iii) when $n_0 < n_2$, $|f(s_1, s_2)|$ is maximized when $|g(s_1, s_2)|$ is maximized or minimized as well. Because

$$g(s_1, s_2) = (s_1 + 2s_2) - \frac{S(n_1 + 2n_2)^2}{2N(n_2 - n_0)}$$

$$= (s_1 + 2s_2) - \frac{S(N + n_2 - n_0)^2}{2N(n_2 - n_0)}$$

$$= (s_1 + 2s_2) - S\left[1 + \frac{N^2 + (n_2 - n_0)^2}{2N(n_2 - n_0)}\right]$$

$$\leq (s_1 + 2s_2) - 2S, \quad \text{because } N^2 + (n_2 - n_0)^2 \geq 2N(n_2 - n_0)$$

$$\leq 0,$$

$|g(s_1, s_2)|$ is maximized when $(s_1 + 2s_2)$ is minimized, and it is minimized when $(s_1 + 2s_2)$ is maximized.

The preceding analysis shows that for any given $n_1$ and $n_2$, $\left|\frac{\partial}{\partial n_1}Y_A(v^s)\right|$ is maximized when $(s_1 + 2s_2)$ is maximized or minimized; in other words, to maximize $\left|\frac{\partial}{\partial n_1}Y_A(v^s)\right|$, we only need to consider tables for which $(s_1, s_2) = (0, 0)$ or $(s_1, s_2) = (n_1, n_2)$.

Given $(s_1, s_2) = (0, 0)$, we have $C = -\frac{S}{N}(n_1 + 2n_2)$, and

$$\frac{\partial}{\partial n_1} Y_A(v^s) \bigg/ (-2\alpha) = (n_0 - n_2)\frac{C^2}{D^2} + \frac{SC}{ND}$$

$$= (n_0 - n_2)\frac{\left[\frac{S}{N}(n_1 + 2n_2)\right]^2}{[(n_1 + 2n_2)(2n_0 + n_1)]^2} + \frac{S}{N}\frac{-\frac{S}{N}(n_1 + 2n_2)}{(n_1 + 2n_2)(2n_0 + n_1)}$$

$$= -\frac{S^2}{N(2n_0 + n_1)^2}.$$

So $\left|\frac{\partial}{\partial n_1} Y_A(v^s)\right|$ is maximized when $2n_0 + n_1$ is minimized. Because $(r_0 > 0, s_1 = s_2 = 0) \implies (r_0 \geq 1, r_1 = n_1, r_2 = n_2, s_0 = S) \implies (n_0 \geq S + 1, n_1 \geq 1)$, the minimum occurs at $v^s = (0, 0, 1, R - 2)$, i.e.,

$$\begin{cases} r_0 = 1, & r_1 = 1, & r_2 = R - 2, \\ s_0 = S, & s_1 = 0, & s_2 = 0, \\ n_0 = S + 1, & n_1 = 1, & n_2 = R - 2. \end{cases}$$

Given $(s_1, s_2) = (n_1, n_2)$, we have $C = \frac{R}{N}(n_1 + 2n_2)$, and

$$\frac{\partial}{\partial n_1} Y_A(v^s) \bigg/ (-2\alpha) = (n_0 - n_2)\frac{\left[\frac{R}{N}(n_1 + 2n_2)\right]^2}{[(n_1 + 2n_2)(2n_0 + n_1)]^2} + \frac{S}{N}\frac{\frac{R}{N}(n_1 + 2n_2)}{(n_1 + 2n_2)(2n_0 + n_1)}$$

$$= \frac{R(S + n_0 - n_2)}{N(2n_0 + n_1)^2}$$

$$= -\frac{1}{N}\left[\left(\frac{R}{2n_0 + n_1} - \frac{1}{2}\right)^2 - \frac{1}{4}\right].$$

Because $(s_1 = n_1, s_2 = n_2) \implies (r_1 = r_2 = 0) \implies (r_0 = R) \implies (n_0 \geq R)$, we have $0 < \frac{R}{2n_0 + n_1} < 1/2 \implies 0 < \left(\frac{R}{2n_0 + n_1} - \frac{1}{2}\right)^2 < 1/4$. So $\left|\frac{\partial}{\partial n_1} Y_A(v^s)\right|$ is maximized when $\left(\frac{R}{2n_0 + n_1} - \frac{1}{2}\right)^2$ is minimized, which is achieved when $2n_0 + n_1$ is minimized, which occurs at $v^s = (1, S - 1, 1, S - 1)$, i.e.,

$$\begin{cases} r_0 = R, & r_1 = 0, & r_2 = 0, \\ s_0 = 0, & s_1 = 1, & s_2 = S - 1, \\ n_0 = R, & n_1 = 1, & n_2 = S - 1. \end{cases}$$

To summarize, when $r_0 > 0$, for any table $v^s$, the change of $Y_A(v^s)$ in the direction of $u_2^s$ is no less than that in the direction of $u_1^s$. The maximum change of $Y_A$ in the direction of $u_2 = (-1, 0, 1, 0, 0, 0, -1, 0, 1) \equiv u_2^s$ occurs

$$
\begin{aligned}
\text{at } v_1^* &= (1, 1, R-2, S, 0, 0, S+1, 1, R-2), \\
\text{with } \Delta_1 &= |Y_A(v_1^*) - Y_A(v_1^* + u_2)| \\
&= \frac{2N^3}{RS} \left(\frac{S}{N}\right)^2 \left| \frac{2R-3}{2N-(2R-3)} - \frac{2R-1}{2N-(2R-1)} \right| \\
&= \frac{8N^2 S}{R(2S+3)(2S+1)}, \\
\text{or at } v_2^* &= (R, 0, 0, 0, 1, S-1, R, 1, S-1), \\
\text{with } \Delta_2 &= |Y_A(v_2^*) - Y_A(v_2^* + u_2)| \\
&= \frac{2N^3}{RS} \left\{ \left(\frac{R}{N}\right)^2 \frac{2S-1}{2R+1} - \frac{[\frac{R}{N}(2S+1) - 2]^2}{(2S+1)(2R-1)} \right\} \\
&= \frac{8N^2 [R^2(2S-1) - S]}{RS(2S+1)(2R+1)(2R-1)}.
\end{aligned}
$$

The same analysis for $s_0 > 0$ reveals that $|\nabla Y_A(v^r) \cdot u_4| = 2|\nabla Y_A(v^r) \cdot u_3| = 2\left|\frac{\partial}{\partial n_1} Y_A(v^r)\right|$, and the maximum change of $Y_A$ in the direction of $u_4 = (0, 0, 0, -1, 0, 1, -1, 0, 1) \equiv u_2^r$ occurs

$$
\begin{aligned}
\text{at } v_3^* &= (R, 0, 0, 1, 1, S-2, R+1, 1, S-2), \\
\text{with } \Delta_3 &= |Y_A(v_3^*) - Y_A(v_3^* + u_4)| = \frac{8N^2 R}{S(2R+3)(2R+1)}, \\
\text{or at } v_4^* &= (0, 1, R-1, S, 0, 0, S, 1, R-1), \\
\text{with } \Delta_4 &= |Y_A(v_4^*) - Y_A(v_4^* + u_4)| = \frac{8N^2 [S^2(2R-1) - R]}{RS(2R+1)(2S+1)(2S-1)}.
\end{aligned}
$$

$\square$

## Appendix C. Proof of Theorem 4.1

*Proof.* To show that Algorithm 2 is $\epsilon$-differentially private, it suffices to show that choosing $\mathcal{S}$ is $\epsilon/2$-differentially private. The rest of the proof follows from the proof of Algorithm 1 in Uhler et al. [2].

Following the notation in McSherry and Talwar [24], we define the random variable of sampling a single SNP, $\varepsilon_q^\epsilon$, by

$$\Pr(\varepsilon_q^\epsilon(D) = i) \propto \exp\left(\frac{\epsilon q(D,i)}{2\Delta_q}\right) \mu(i)$$

$$\propto \exp\left(\frac{\epsilon q(D,i)}{2s}\right)$$

where $q(D,i)$ is the score for $\text{SNP}_i$, $s$ is the sensitivity for the scoring function $q(D,i)$, and $\mu(i) = 1/M'$ is constant. We also define

$$q_B(D,i) = \begin{cases} \text{score of the } \text{SNP}_i & \text{if } i \notin B \\ -\infty & \text{if } i \in B \end{cases}.$$

where $B$ is a set of SNPs and $q_B$ denotes the scoring function given that the SNPs in $B$ have been sampled and thus have 0 sampling probability in subsequent sampling steps. Note that

$$\Pr(\varepsilon_{q_B}^\epsilon(D) = i, i \notin B) = \frac{\exp\left(\frac{\epsilon q_B(D,i)}{2s}\right)}{\sum_{j \notin B} \exp\left(\frac{\epsilon q_B(D,j)}{2s}\right)}$$

$$\leq \frac{\exp\left(\frac{\epsilon [q_B(D',r)+s]}{2s}\right)}{\sum_{j \notin B} \exp\left(\frac{\epsilon [q_B(D',r)-s]}{2s}\right)}$$

$$= e^\epsilon \ \Pr(\varepsilon_{q_B}^\epsilon(D') = i, i \notin B).$$

Let $\sigma$ denote a permutation of $\mathcal{S}$.

$$\Pr(\text{sampling } \mathcal{S}|D) = \sum_{\sigma \in \sigma(S)} \Pr(\varepsilon_q^{\epsilon/(2M)}(D) = \sigma(1)) \prod_{i=2}^{M} \Pr(\varepsilon_{q_{\{\sigma(j), j<i\}}}^{\epsilon/(2M)}(D) = \sigma(i))$$

$$\leq \sum_{\sigma \in \sigma(S)} \left\{ e^{\epsilon/(2M)} \ \Pr(\varepsilon_q^{\epsilon/(2M)}(D') = \sigma(1)) \right\}$$

$$\prod_{i=2}^{M} \left\{ e^{\epsilon/(2M)} \ \Pr(\varepsilon_{q_{\{\sigma(j), j<i\}}}^{\epsilon/(2M)}(D') = \sigma(i)) \right\}$$

$$= e^{\epsilon/2} \Pr(\text{sampling } \mathcal{S}|D').$$

$\square$